\documentclass[doublespacing]{elsart}

\usepackage{epsfig}
\usepackage{graphicx}

\usepackage{amssymb}

\journal{Solid State Electronics}

\begin{document}

\begin{frontmatter}


\title{Growth of Single Unit-Cell Superconducting La$_{2-x}$Sr$_x$CuO$_{4}$ Films}
\author[neuch,ibm]{A. R\"ufenacht\corauthref{cor1}},
\author[neuch]{P. Chappatte},
\author[geneve,ibm]{S. Gariglio},
\author[neuch]{C. Leemann},
\author[ibm]{J. Fompeyrine},
\author[ibm]{J.-P. Locquet}, and
\author[neuch]{P. Martinoli}

\corauth[cor1]{Corresponding author : alain.rufenacht@unine.ch -
FAX : +41 32 718 29 01}
\address[neuch]{Institut de Physique, Universit\'e de Neuch\^atel, CH-2000 Neuch\^atel}
\address[ibm]{IBM Research Division, Zurich Research Laboratory, CH-8803 R\"uschlikon}
\address[geneve]{DPMC, Universit\'e de Gen\`eve, CH-1211 Gen\`eve 4}

\begin{abstract}

We have developed an approach to grow high quality ultrathin films
of  La$_{2-x}$Sr$_x$CuO$_{4}$ with molecular beam epitaxy, by
adding a homoepitaxial buffer layer in order to minimize the
degradation of the film structure at the interface. The advantage
of this method is to enable a further reduction of the minimal
thickness of a superconducting La$_{1.9}$Sr$_{0.1}$CuO$_{4}$ film.
The main result of our work is that a single unit cell (only two
copper oxide planes) grown on a SrLaAlO$_4$ substrate exhibits a
superconducting transition at 12.5 K (zero resistance) and an
in-plane magnetic penetration depth $\lambda_{ab}(0)$ = 535 nm.

\end{abstract}

\end{frontmatter}

\section{Introduction}

Growing very thin superconducting films with well-characterized
and interesting physical properties is a difficult challenge for
the material scientist. In an attempt to reduce the smallest
superconducting thickness, we have developed a homoepitaxial
technique for the growth of La$_{2-x}$Sr$_x$CuO$_{4}$ (LSCO),
using a normal metallic buffer layer with a strontium (Sr)
concentration x=0.4. On this metallic buffer, we grow the
physically relevant structure, i.e. a few unit cells of
La$_{1.9}$Sr$_{0.1}$CuO$_4$. On top of this film an additional two
unit cell thick La$_{1.6}$Sr$_{0.4}$CuO$_4$ metallic cap (x=0.4)
is deposed for protection (Fig.~\ref{fig1} a).

So far the thinnest superconducting optimally doped LSCO film
directly grown on a SrLaAlO$_4$ substrate was 2 unit
cell\footnote{Notice that the crystallographic unit cell of LSCO
contains two copper oxide planes.} (UC) thick with a
superconducting critical temperature of 10 K and a
superconductivity onset around 35 K \cite{PhysicaC_274_96}. We
have reproduced this result, with the same thickness on the same
substrate, resulting in a T$_c$(R=0) of 10 K and a
superconductivity onset near 35 K (SLAO 0/2/0).

The use of our buffer technique eliminates the hazardous effects
of the film-susbtrate interface (due to mismatch and chemical
interactions) on the superconducting properties of these extremely
thin films. Our method allows the experimentalist to grow quasi
two-dimensional superconducting films, an interesting tool for
physical investigations and device applications.

\section{Experimental and fabrication details}

The samples studied in this report were thin films grown in a
molecular beam epitaxy (MBE) system. The sample fabrication
details were reported elsewhere \cite{spie94}. The method used to
grow epitaxial thin films consists of a block-by-block deposition
\cite{MRS94}. The oxygenation of the films during the deposition
and the cool down is provided by a RF plasma source of atomic
oxygen. A reflection high-energy electron diffraction (RHEED)
system allows us to monitor the growth process, in particular the
quality of the in-plane growth.

\begin{figure}[T]
\begin{center}

\centering{
\includegraphics[width=11cm]{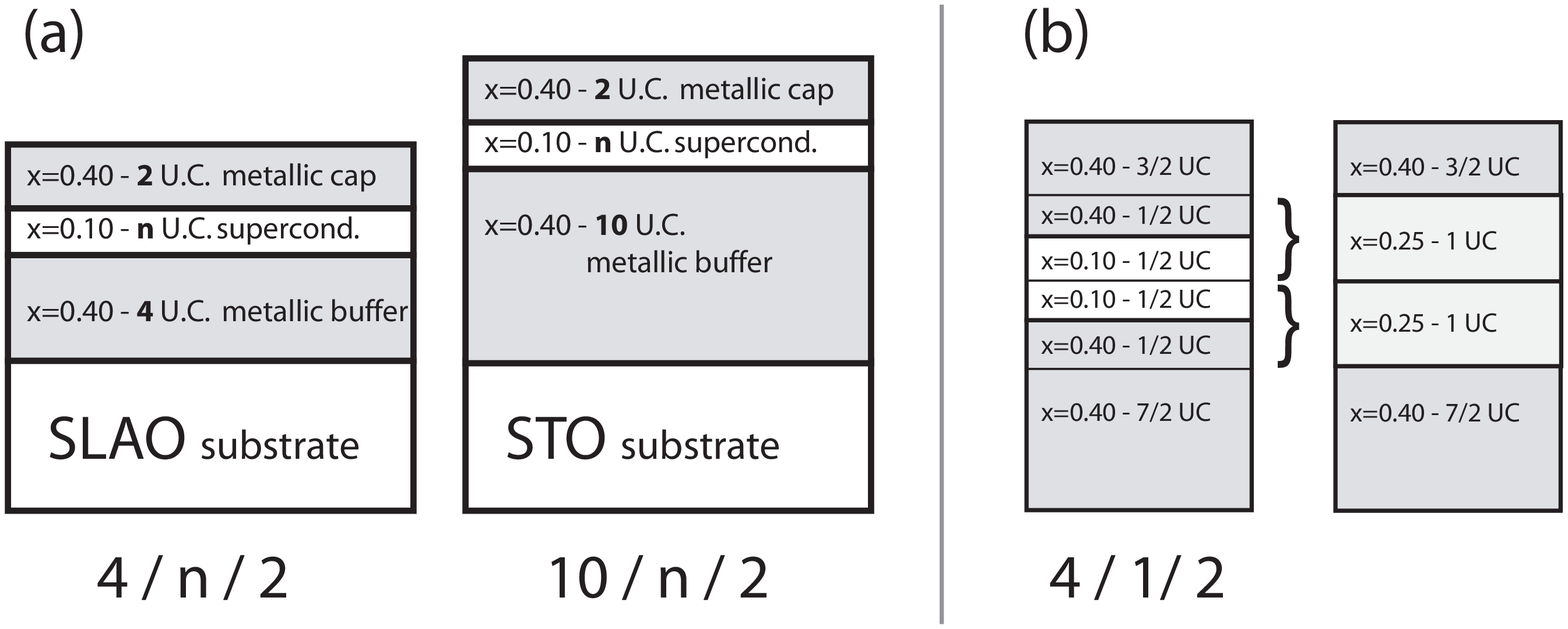}}
\caption[fig1]{\small Homoepitaxial buffer technique applied to
La$_{2-x}$Sr$_x$CuO$_{4}$ compounds. (\textbf{a}): general scheme
for the two different kinds of substrate. Notation: B/F/C
corresponds, respectively, to the number of buffer layer unit
cells (B), superconducting thin film (F), and metallic cap (C).
(\textbf{b}): model for the influence of Sr interdiffusion in the
SLAO 4/1/2 compound.} \label{fig1}
\end{center}
\end{figure}

The choice of the substrate and the resulting strain effects play
an important role in the structural quality \cite{spie98} and
physical properties \cite{nature98} of the film. In this work,
ultra-thin films were deposited on ($100$) SrTiO$_3$ (STO) and
($001$) SrLaAlO$_4$ (SLAO) substrates. The thickness of the buffer
layers was adapted to the two substrates. Due to a larger
mismatch, the buffer layer on a STO substrate is thicker (10 UC)
than on a SLAO substrate (4 UC), allowing a better film-substrate
interface decoupling\footnote{SLAO (compressive) $a_{axis}=3.754$
\AA, STO (tensile) $a_{axis}=3.905$ \AA, LSCO with $x=0.1$,
$a_{axis}=3.784$ \AA.} (Fig.~\ref{fig1} a).

In order to understand whether the Sr interlayer diffusion across
the interface increases the effective number of UC necessary for
superconductivity, we consider the case where the interlayer
diffusion is limited to the single unit cell film with nominal
concentration x=0.1 and the adjacent bottom and top layers,
half-UC thick, with nominal concentration x=0.4. The rest of the
layers are assumed to act as reservoirs with an almost constant
concentration x=0.4. The interaction of the upper superconducting
half-UC with the first cap half-UC will give for each half-UC an
average concentration $x=(0.1+0.4)/2=0.25$ (Fig.~\ref{fig1} b).
The scenario for the lower superconducting half-UC is identical.
The final result would be a film with 4 half UC with x=0.25. LSCO
compounds with this Sr concentration are metallic and at the upper
overdoped limit of the superconducting phase boundary. With this
assumption, a lower amount of Sr diffusion will not increase the
number of superconducting layers. However, we can not exclude a
rise of the effective doping level in the superconducting
layer(s).

Instead of using a metallic buffer, it is also possible to choose
an insulating material, by reducing the Sr content of the LSCO
buffer below x=0.05. In this way, the Sr in-diffusion is excluded.
However, such LCO and low doped LSCO can be driven superconducting
by adding interstitial oxygen \cite{PRB54_96}. This process occurs
preferentially at low Sr concentration. Note that the growing LSCO
or LCO films doped with interstitial oxygen is more difficult to
master than by varying the Sr concentration.

\section{Structural properties}

\begin{figure}[H]
\begin{center}

\centering{
\includegraphics[width=11cm]{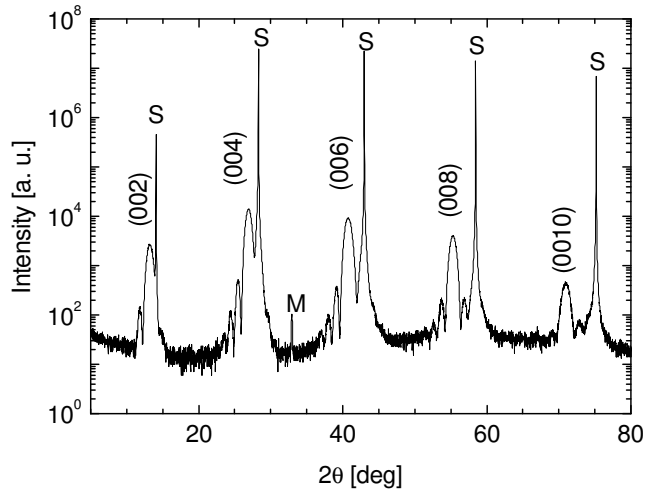}}
\caption[fig2]{\small X-ray
diffraction pattern of 4/1/2 LSCO thin film on SLAO:
$\theta$-2$\theta$ scan showing the LSCO (00$\ell$) reflection
with their finite size effect oscillations as well as the peaks
due to the substrate (S). The peak denoted by (M) comes from the
material used to hold the sample in the diffractometer. }
\label{fig2}
\end{center}
\end{figure}

The x-ray diffraction measurements confirm the $c$-axis growth of
the films deposited on SLAO and STO. From conventional
$\theta-2\theta$ scans, a linear regression of the (00$\ell$)
peaks positions gives an average lattice parameter\footnote{to be
compared with the bulk value (x=0.4) : $c$=(13.26$\pm$0.01) \AA~
\cite{PRB49_94}} which for the film 4/1/2 on SLAO substrate is
$c$=(13.285$\pm$0.010) \AA~(Fig.~\ref{fig2}). Finite-size effect
oscillations were observed at low angle and around the (00$\ell$)
peaks ($\ell$ = 2, 4, 6, 8), allowing a good estimate of the total
film thickness. For the 4/1/2 film grown on SLAO, finite-size
effect oscillation observed around the (004) peak give a thickness
of 89.4 \AA , in agreement with the presence of 7 UC of 13.285
\AA. Pole figure of the (103) plane of the film, measured on the
same sample, confirms the epitaxial growth.

\section{Superconducting properties}
\subsection{Resistivity measurements}

\begin{figure}[H]
\begin{center}

\centering{
\includegraphics[width=11cm]{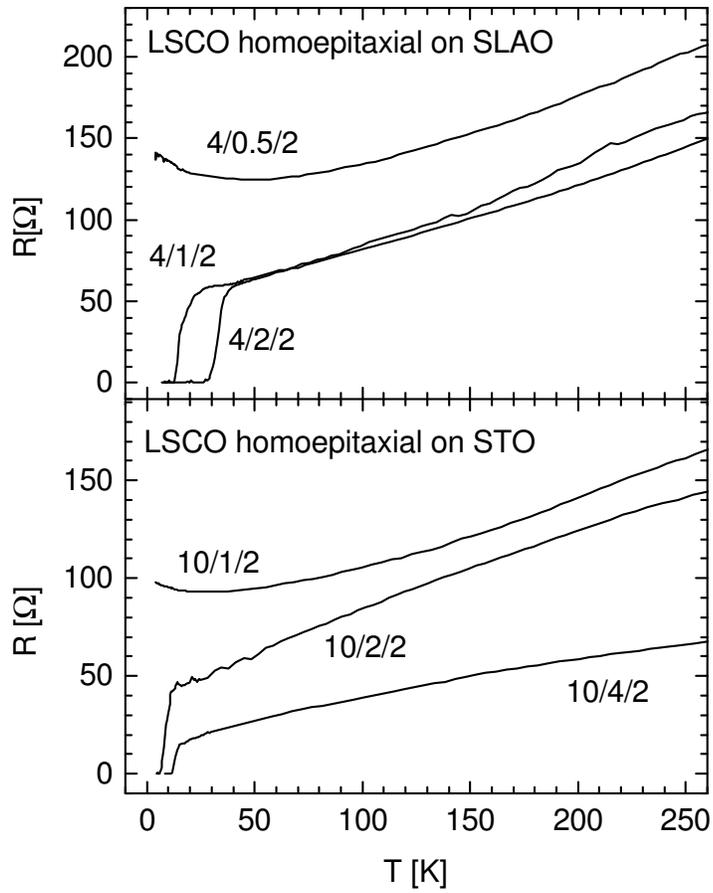}}
\caption[fig3]{\small Resistance measurements for different
thicknesses and kinds of substrate. The curves are plotted in
terms of the resistance as measured between two indium contacts
(10 mm long) separated by 4 mm.} \label{fig3}
\end{center}
\end{figure}

Resistivity measurements were performed in a cryoprobe (range 300K
$\rightarrow$ 4.2 K), using a standard DC four-point method.
Contacts were made with four indium wires pressed onto the top of
the film structures. The measurements (Fig.~\ref{fig3}) performed
on different samples provide evidence of superconductivity in
samples as thin as 1 UC on SLAO and 2 UC on STO. No
superconductivity was observed in the film with a similar nominal
Sr concentration but with only one copper oxide plane. These
results demonstrate the validity of the homoeptiaxial buffer layer
technique and tell us that the Sr interdiffusion is small enough
not to depress the superconducting behavior. The thickness has a
major influence on the superconducting properties. The
superconducting critical temperatures T$_{c0}$ (as defined by the
bottom of the resistivity transition) for the layers grown on SLAO
are respectively 12.5 K and 25 K for the 1 UC and 2 UC thin films.
On STO substrates, the T$_{c0}$ are 5.7 K for a 2 UC and 11.3 K
for a 4 UC thin film. The fact that T$_{c0}$ is found to be
proportional to the film thickness is consistent with
two-dimensional behavior \cite{Schneider}. At identical thickness,
the critical temperature of layers grown on STO is much lower than
the one obtained for a film grown on the SLAO. The overall
resistance of the whole three-layers structure results from the
contributions of the individual layers, which have not been
investigated in detail. Nevertheless, the temperature dependence
in the normal state is consistent with the value obtained for
x=0.4 doped LSCO films \cite{PRB61_00}.

\subsection{Magnetic penetration depth measurements}
The measurements of the magnetic penetration depth were performed
using a two-coil technique \cite{APL55_89} . Briefly, this method
measures, contact-less and with an astatically wound coil, the
response (due to screening currents) of a sample excited by an
external AC electromagnetic field. The complex sheet impedance of
the film is extracted from the signal measured with a lock-in
amplifier. The inductive part of the sheet impedance of the
superconducting film (the so called kinetic inductance $L_k$) is
related to the bulk in-plane magnetic penetration depth
$\lambda_{ab}$ by the following relation: $
L_k(T)=\frac{\mu_0\lambda_{ab}^2(T)}{d}$, where $d$ is the
thickness of the film.

All inductive measurements were performed in a $^3$He cryostat.
Notice that the inductive method probes the onset of global
superconducting phase coherence. For this reason the T$_c$ deduced
from the inductive measurements is lower than that extracted from
the resistive transitions \cite{PRL64_90}.

\begin{figure}[H]
\begin{center}

\centering{
\includegraphics[width=11cm]{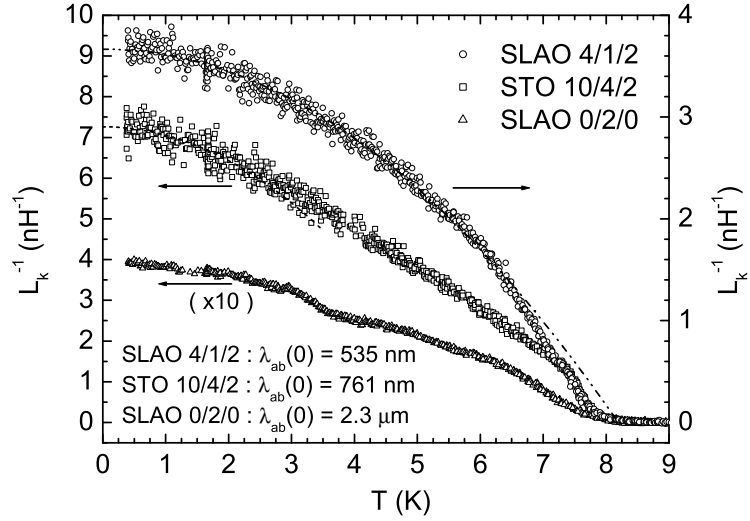}}\caption[fig4]{\small Inverse kinetic inductance ($\propto
\lambda_{ab}^{-2}$) vs the temperature for samples on different
substrates and with different thicknesses. The inductive
measurement presented here was performed with a frequency of 1kHz
and a drive current of 10 $\mu$A. All films (LSCO 4/1/2, STO
10/4/2 and buffer-less SLAO 0/2/0) present a similar critical
temperature. The dashed lines show the quadratic fit of the data
in the low temperature regime.} \label{fig4}
\end{center}
\end{figure}

As shown in figure \ref{fig4}, the zero-temperature value
$\lambda_{ab}^{-2}(0)$ for the 1 UC film grown on SLAO was deduced
by fitting the kinetic inductance data to the parabolic expression
$\lambda_{ab}^{-2}(T)= \lambda_{ab}^{-2}(0)\cdot
(1-(\frac{T}{Tc})^2)$ \cite{Tinkham}. A similar procedure was used
to fit the low-temperature data of the 4 UC film grown on STO.
Although both films exhibit almost identical critical temperatures
($\sim$8.5 K), the 1 UC film has a smaller value of the magnetic
penetration depth ($\lambda_{ab}(0)$=535 nm) than the 4 UC film
($\lambda_{ab}(0)$=760 nm). This shows the superior structural
quality of the film grown on SLAO, in agreement with previous
observations \cite{nature98}. We also notice that the values of
$\lambda_{ab}(0)$ obtained using the buffer technique are much
smaller than that ($\lambda_{ab}(0)$= 2.3 $\mu$m) for a 2 UC film
grown directly on SLAO (see Fig. \ref{fig4}), thereby
demonstrating the significant advantage of our method. An other
remarkable feature emerges if one compares the penetration depth
of the 1 UC film grown on SLAO ($\lambda_{ab}(0)$=535 nm) with
that of optimally doped LSCO (bulk) single crystals (Tc=35 K and
$\lambda_{ab}(0)$=300 nm \cite{PRL72_94}). Assuming
two-dimensional behavior, for which one expects
$T_c\propto\lambda_{ab}^{-2}$, one deduces that the structural
quality of the ultrathin films grown on SLAO with our
homoepitaxial buffer technique is comparable to that of LSCO
single crystals.

\section{Conclusions}

The homoepitaxial buffer technique is an efficient tool to further
reduce the superconducting minimal thickness. An X-ray
characterization proved the high structural quality (c-axis
orientation and epitaxy) of the films. Values obtained for the
penetration depth and T$_c$ confirm the correlated role of
substrate choice and homoepitaxial buffer technique when growing
very high quality ultra-thin La$_{2-x}$Sr$_x$CuO$_{4}$ films.
Finally, the work we present here forms a new starting point for
investigations of physical properties in ultra-thin
superconducting layers and research of device applications.

\section{Acknowledgments}

We thank M. Dodgson, H. Siegwart, C. Rossel, A. Guiller, J. W.
Seo, E. Koller and J.-M. Triscone for helpful discussions. This
work was supported by the Swiss National Science Foundation.




\clearpage

\end{document}